\documentclass[10pt]{iopart}
\usepackage{iopams,setstack}
\usepackage{amssymb,graphicx,subfig,color}
\usepackage[pdfborder={0 0 0}]{hyperref}

\begin{document}

\newcommand{\pa}{\partial}
\newcommand{\eps}{\epsilon}
\newcommand{\DW}{\Delta W}
\newcommand{\DF}{\Delta F}
\newcommand{\Dk}{\Delta}
\newcommand{\cd}{{\!\!\!\!\!\!\!\!\cal D}x(\cdot)}
\newcommand{\dotx}{\dot{x}}
\newcommand{\doty}{\dot{y}}
\newcommand{\dV}{\dot{V}}
\newcommand{\tL}{\tilde{L}}
\newcommand{\la}{\langle}
\newcommand{\ra}{\rangle}
\newcommand{\bx}{\bar{x}}
\newcommand{\lla}{\left\langle}
\newcommand{\rra}{\right\rangle}
\newcommand{\llbr}{\left\lbrace}
\newcommand{\rrbr}{\right\rbrace}
\newcommand{\bxt}{\dot{\bar{x}}}
\newcommand{\bq}{\bar{q}}
\newcommand{\bS}{\bar{S}}
\newcommand{\bR}{\bar{R}}
\newcommand{\bV}{\bar{V}}
\newcommand{\bdV}{\dot{\bar{V}}}
\newcommand{\N}{{\cal N}}
\newcommand{\ph}{\varphi}
\newcommand{\dph}{\dot{\varphi}}
\newcommand{\J}{{\cal J}}
\newcommand{\lam}{\lambda}
\newcommand{\al}{\alpha}
\newcommand{\di}{\!\!\!\mathrm{d}}
\renewcommand{\d}{\mathrm{d}}
\renewcommand{\i}{\mathrm{i}}
\renewcommand{\e}{\mathrm{e}}
\newcommand{\nn}{\nonumber}

\title{Asymptotic work distributions in driven bistable systems}%
\author{D. Nickelsen$^1$, A. Engel$^1$}
\address{$^1$ Universit\"{a}t Oldenburg, Institut f\"{u}r Physik, 26111
  Oldenburg, Germany}

\eads{\mailto{daniel.nickelsen@uni-oldenburg.de}, \mailto{andreas.engel@uni-oldenburg.de}}
%\thanks{}%
%\subjclass{}%
%\keywords{}%

%\date{\today}
%\dedicatory{}%

% ----------------------------------------------------------------
\begin{abstract}
The asymptotic tails of the probability distributions of thermodynamic quantities convey important information about the physics of nanoscopic systems driven out of equilibrium. We apply a recently proposed method to analytically determine the asymptotics of work distributions in Langevin systems to an one-dimensional model of single-molecule force spectroscopy. The results are in excellent agreement with numerical simulations, even in the centre of the distributions. We compare our findings with a recent proposal for an universal form of the asymptotics of work distributions in single-molecule experiments. 
\end{abstract}

\pacs{05.70.Ln, 02.50.-r, 05.40.-a, 05.20.-y}

\maketitle

%%%%%%%%%%%%%%%%%%%%%%%%%%%%%%%%%%%%%%%%%%%%%%%%%%%%%%%%%%%%%%%%%%%%%%%%%%%%%%%%%%%%%%%%%%%%%%%%%%%%%%
%%%%%%%%%%%%%%%%%%%%%%%%%%%%%%%%%%%%%%%%%%%%%%%%%%%%%%%%%%%%%%%%%%%%%%%%%%%%%%%%%%%%%%%%%%%%%%%%%%%%%%

\section{Introduction}
Traditionally, statistical mechanics is concerned with {\em averages}; the probability distributions for thermodynamic quantities of macroscopic systems are so exceedingly sharp that only their most probable values matter. These in turn are practically identical with the averages. Only near instabilities, deviations from averages become important. In static cases these fluctuations are well described by the second moments of the respective distributions; if the dynamics is of importance as well, they are characterized by the correlation functions. 

When investigating nanoscopic systems from the point of view of thermodynamics, the situation changes. Fluctuations are now strong and ubiquitous, and correspondingly, the probability distributions of relevant quantities are broad and poorly characterized by their leading moments alone. Whereas this seems rather obvious, it came as a real surprise that work \cite{Jar} and fluctuation \cite{ECM,GaCo} theorems which form the cornerstones of the emerging field of {\em stochastic thermodynamics} {\cite{Jarrev,Seifert,EsvB,book_ken,Udo_rev} are relations that probe the {\em very far tails} of the respective probability distributions. Very unlikely events now carry significant information about the physics of the system under consideration. On the one hand, interest in mathematical investigations like large deviation theory (see \cite{Touchette} and references therein) is renewed, on the other hand, techniques are in demand that allow to determine the {\em asymptotics} of probability distributions. Since rare events are hard to get in experiments and numerical simulations, approximate analytical procedures have to be developed.

In the present paper we apply a recently proposed method for the analytical determination of the asymptotics of work distributions in driven Langevin systems \cite{Nickelsen2011} to a simple model for single-molecule force spectroscopy. In these experiments (for a recent review see \cite{Ritort_rev}) the free-energy difference $\Delta F$ between the folded and the unfolded state of a biomolecule is determined from the distribution of work $W$ obtained in isothermal unfolding and refolding processes. If only one transition is monitored \cite{Liphardt}, the Jarzynski equation \cite{Jar} 
\begin{eqnarray}\label{eq:JE}
 \lla \e^{-\beta W} \rra = \e^{-\beta \Delta F}
\end{eqnarray} 
is employed, where $\beta$ denotes the inverse of the temperature. If histograms of work for both the forward and the reverse process are compiled \cite{Collin}, it may be advantageous to use the Crooks fluctuation theorem \cite{Crooks}
\begin{eqnarray}\label{eq:CR}
 \frac{P(W)}{P_r(-W)}=\e^{\beta(W-\Delta F)}\; .
\end{eqnarray}
Here, the free-energy difference $\Delta F$ is identified from the intersection of the probability density functions $P(W)$ of the forward and $P_r(-W)$ of the reverse process. Note that in both cases an accurate estimate for $\Delta F$ requires reliable information about the tails of the work probability distributions.

The paper is organized as follows. In section 2 we introduce the notation and review the general method \cite{Nickelsen2011}. Section 3 contains the application to an one-dimensional stochastic process with a time-dependent double-well potential and the comparison with numerical simulations of the system. In section 4 we discuss our results and in particular compare them with a recent proposal for the general shape of the tail of the work distribution in single-molecule experiments \cite{Palassini2011}.

%%%%%%%%%%%%%%%%%%%%%%%%%%%%%%%%%%%%%%%%%%%%%%%%%%%%%%%%%%%%%%%%%%%%%%%%%%%%%%%%%%%%%%%%%%%%%%%%%%%%%%
%%%%%%%%%%%%%%%%%%%%%%%%%%%%%%%%%%%%%%%%%%%%%%%%%%%%%%%%%%%%%%%%%%%%%%%%%%%%%%%%%%%%%%%%%%%%%%%%%%%%%%

\section{General Theory}
To pave the way for the analysis and to fix the notation we summarize in the present section in a very condensed way the main steps of our approach to determine the asymptotic tail of work distributions in driven Langevin systems. For more details the reader is referred to  \cite{Nickelsen2011}. 

We investigate a system described by an one-dimensional, overdamped Langevin equation which in dimensionless units has the form 
\begin{eqnarray}
  \label{eq:LE}
  \dot{x}=-V'(x,t) +\sqrt{2/\beta}\; \xi(t)\; .
\end{eqnarray}
Here $x$ denotes the degree of freedom, $V$ is a time-dependent potential modelling the external driving, and $\xi(t)$ is Gaussian white noise with 
$\la\xi(t)\ra\equiv 0$ and $\la \xi(t)\xi(t')\ra=\delta(t-t')$. We denote derivatives with respect to $x$ by a prime, and those with respect to $t$ by a dot. The initial state $x(t=0)=:x_0$ of the process is sampled from the equilibrium distribution at $t=0$ with  inverse temperature $\beta$,
\begin{eqnarray}\label{eq:rho_0}
  \rho_0(x_0)=\frac{1}{Z_0}\exp\left[-\beta V_0(x_0)\right]\; .
\end{eqnarray}
Accordingly, $V_0(x):=V(x,t=0)$ is the initial potential and 
\begin{eqnarray}\label{eq:defZ_0}
  Z_0=\int\! \d x \exp\left[-\beta V_0(x)\right]
\end{eqnarray} 
the corresponding partition function. 

The work performed on the system for a particular trajectory $x(\cdot)$ is given by \cite{Sekimoto} 
\begin{eqnarray}
  \label{eq:defwork}
  W[x(\cdot)]=\int_0^{T} \di t\; \dot{V}\big(x(t),t\big)\; .
\end{eqnarray}
Due to the randomness inherent in $x(\cdot)$, the work $W$ is itself a random quantity and its pdf can be written as 
\begin{eqnarray}
  \label{eq:defPofW} 
  \eqalign{
  \fl P(W)=\int\! \frac{\d x_0}{Z_0}\; \exp\left[-\beta V_0(x_0)\right]  \int\! \d x_T \\
       \int\limits_{x(0)=x_0}^{x(T)=x_T} \cd\; p[x(\cdot)] \; 
       \delta(W- W[x(\cdot)]) \;. }
\end{eqnarray}
For mid-point discretization we have \cite{ChDe} 
\begin{eqnarray}\label{eq:defp}
  \fl p[x(\cdot)]=\N[x(\cdot)]\; \exp\left[-\frac{\beta}{4}\int_0^{T} \di t\; \big(\dotx+V'(x,t)\big)^2\right]\; 
\end{eqnarray}
with the normalization factor 
\begin{eqnarray}\label{eq:defN}
 \N[x(\cdot)]=\exp\left[\frac{1}{2}\int_0^T \!\!\!\d t\; V''(x(t),t)\right] \;.
\end{eqnarray}
Hence
\begin{eqnarray}
  \label{eq:PofW}
  \eqalign{
    \fl P(W)=\int\! \frac{\d x_0}{Z_0}\int\! \d x_T \int\! \frac{\d q}{4\pi/\beta} \\
    \int\limits_{x(0)=x_0}^{x(T)=x_T}\cd \,\N[x(\cdot)] \exp\llbr-\beta S[x(\cdot),q]\rrbr }
\end{eqnarray}
with the action 
\begin{eqnarray}
  \label{eq:defS}
  \eqalign{
  \fl S[x(\cdot),q] \!=\! V_0(x_0)\!+\!\int_0^{T}\!\di t 
    \left[\frac{1}{4}(\dot{x}\!+\!V')^2\!+\!\frac{\i q}{2}\dot{V}\right]\!-\!\frac{\i q}{2}W \,. }
\end{eqnarray}
We are interested in the asymptotic behaviour of $P(W)$. Rare values of $W$ are brought about by unlikely trajectories $x(\cdot)$. In the spirit of the contraction principle of large deviation theory \cite{Touchette}, we expect that in the asymptotic tails of $P(W)$ there is {\em one} trajectory for each value of $W$ that dominates $P(W)$. To find it, we have to maximize  $P[x(\cdot)]:=\rho_0(x_0)p[x(\cdot)]$ under the constraint ${W=W[x(\cdot)]}$. This can be done by using a saddle-point approximation of the integrals in (\ref{eq:PofW}). The result is
\begin{eqnarray}\label{eq:main}
 \fl P(W)=\frac{\bar{\N}\sqrt{2}}{Z_0}\,\frac{\exp\left[-\beta\bS\right]}{\sqrt{\bR\,\det A}}
      \big(1+\Or(1/\beta)\big)\, .
\end{eqnarray}
To use this expression in explicit examples, we first have to determine the most probable trajectory $\bx(\cdot)$ satisfying the Euler-Lagrange equation (ELE) 
\begin{eqnarray}\label{eq:ELE}
 \ddot{\bx}+(1-\i\bq)\bdV'-\bV'\bV''=0
\end{eqnarray}
where $\bV(t):=V(\bx(t),t)$ and similarly for derivatives of $V$. The ELE is completed by the 
boundary conditions
\begin{eqnarray}\label{eq:elebc}
 \dot{\bx}_0-\bV'_0=0, \qquad \dot{\bx}_T+\bV'_T=0
\end{eqnarray}
and by the corresponding value $\bq$ of the Lagrange parameter $q$ ensuring $W[\bx(\cdot)]=W$. Using $\bx(\cdot)$ and $\bq$, we calculate $\bS:=S[\bx(\cdot),\bq]$ and $\bar{\N}:=\N[\bx(\cdot)]$. Then all terms in (\ref{eq:main}) depending {\em solely} on the optimal trajectory are determined.

The denominator $\sqrt{\bR\,\det A}$ in (\ref{eq:main}) comprises the contribution from the {\em neighbourhood} of the optimal path and stems from the integral over the Gaussian fluctuations around $\bx(\cdot)$ and $\bq$. Here, 
\begin{eqnarray}\label{eq:defA}
 \fl A:= -\frac{\d^2}{\d t^2}+(\bV'')^2+\bV'\bV'''-(1-\i\bq)\bdV''
\end{eqnarray}
denotes the operator of quadratic fluctuations which acts on functions $\ph(t)$ on the interval $0<t<T$ obeying the boundary conditions
\begin{eqnarray}\label{eq:bcA}
  \eqalign{
    \bV_0''\ph(0)-\dph(0)=0 \;, \\ 
    \bV_T''\ph(T)+\dph(T)=0 \;. 
    }
\end{eqnarray}
A simple prescription to calculate $\det A$ is as follows \cite{KiMcKa}.  
Solve the initial value problem 
\begin{eqnarray}\label{eq:odedetA}
  \eqalign{
    A \chi(t)=0 \;, \\ 
    \chi(0)=1 \;, \qquad \dot{\chi}(0)=\bV_0'' }
\end{eqnarray}
then 
\begin{eqnarray}\label{eq:defF}
 \det A=2\big(\bV_T''\chi(T)+\dot{\chi}(T)\big) \; .
\end{eqnarray}

The factor $\bR$ in (\ref{eq:main}) accounts for the influence of the constraint (\ref{eq:defwork}) on the fluctuations around $\bx(\cdot)$. Since also the trajectories from the neighbourhood of $\bx(\cdot)$ have to yield the very same value of $W$, fluctuations violating this constraint are suppressed. This gives rise to a correction to the fluctuation determinant of the form
\begin{eqnarray}
 \bR=\int_0^T dt\; \bdV'(t) \; A^{-1} \bdV'(t) \;,
\end{eqnarray}
where $A^{-1}$ denotes the inverse operator of $A$. To explicitly determine $\bR$, it is  convenient to solve the ordinary differential equation
\begin{eqnarray}\label{eq:hode}
 A\, \psi(t)=\dV'(\bx(t),t)
\end{eqnarray}
with boundary conditions (\ref{eq:bcA}) and to use 
\begin{eqnarray}\label{eq:resAm1}
 \bR=\int_0^T \!\!\!\d t\; \psi(t) \bdV'(t)\; .
\end{eqnarray}

%%%%%%%%%%%%%%%%%%%%%%%%%%%%%%%%%%%%%%%%%%%%%%%%%%%%%%%%%%%%%%%%%%%%%%%%%%%%%%%%%%%%%%%%%%%%%%%%%%%%%%
%%%%%%%%%%%%%%%%%%%%%%%%%%%%%%%%%%%%%%%%%%%%%%%%%%%%%%%%%%%%%%%%%%%%%%%%%%%%%%%%%%%%%%%%%%%%%%%%%%%%%%

\section{The driven double-well}
The unfolding and refolding of single molecules can be modeled by a time-dependent double-well potential of the form
\begin{eqnarray}\label{e:tilsun_V}
  V(x,t) = ax^4 - bx^2 + r(c-t)x \;,
\end{eqnarray}
where $x$ denotes the extension of the molecule in the direction of the force \cite{Liphardt,Collin,Palassini2011}. The parameters $a$ and $b$ characterize depth and separation of the two minima of $V$, $c$ fixes the moment at which $V$ is symmetric, $V(x)=V(-x)$, and $r$ denotes the transition rate. Choosing $T>c$ for the final time, the two minima will interchange global stability during the process. We have used two exemplary sets of parameters for which the time evolution of $V(x,t)$ is sketched in figure \ref{f:pots}. The main differences between the two sets are that in (b) the minima are further apart and the transition rate $r$ is smaller. 
\begin{figure*} 
  \subfloat[][]{\label{sf:pot1}
	\includegraphics[trim = 0pt 0pt 0pt 30pt, clip, width=0.45\textwidth]{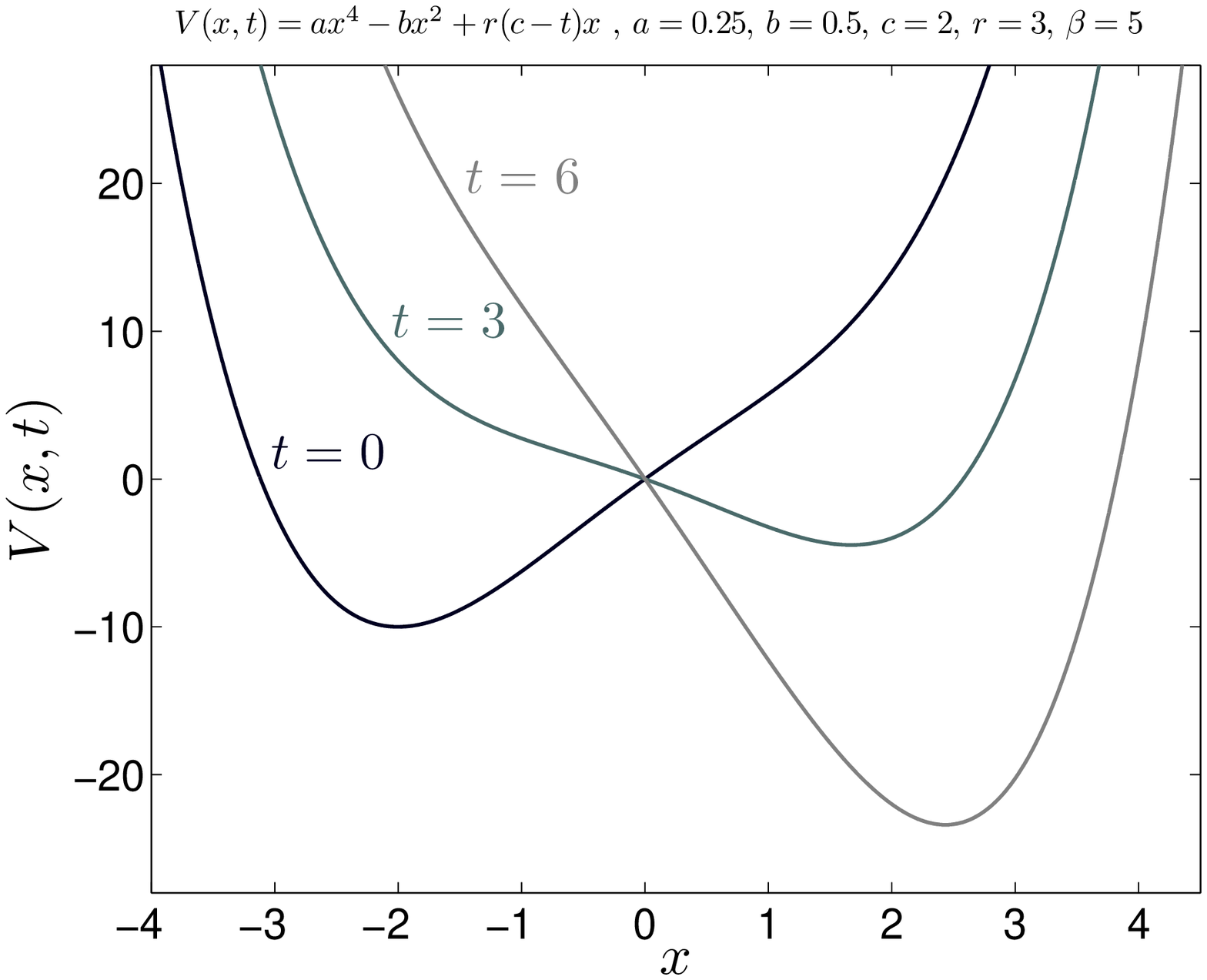}} \qquad % 30
  \subfloat[][]{\label{sf:pot2}
	\includegraphics[trim = 0pt 0pt 0pt 30pt, clip, width=0.45\textwidth]{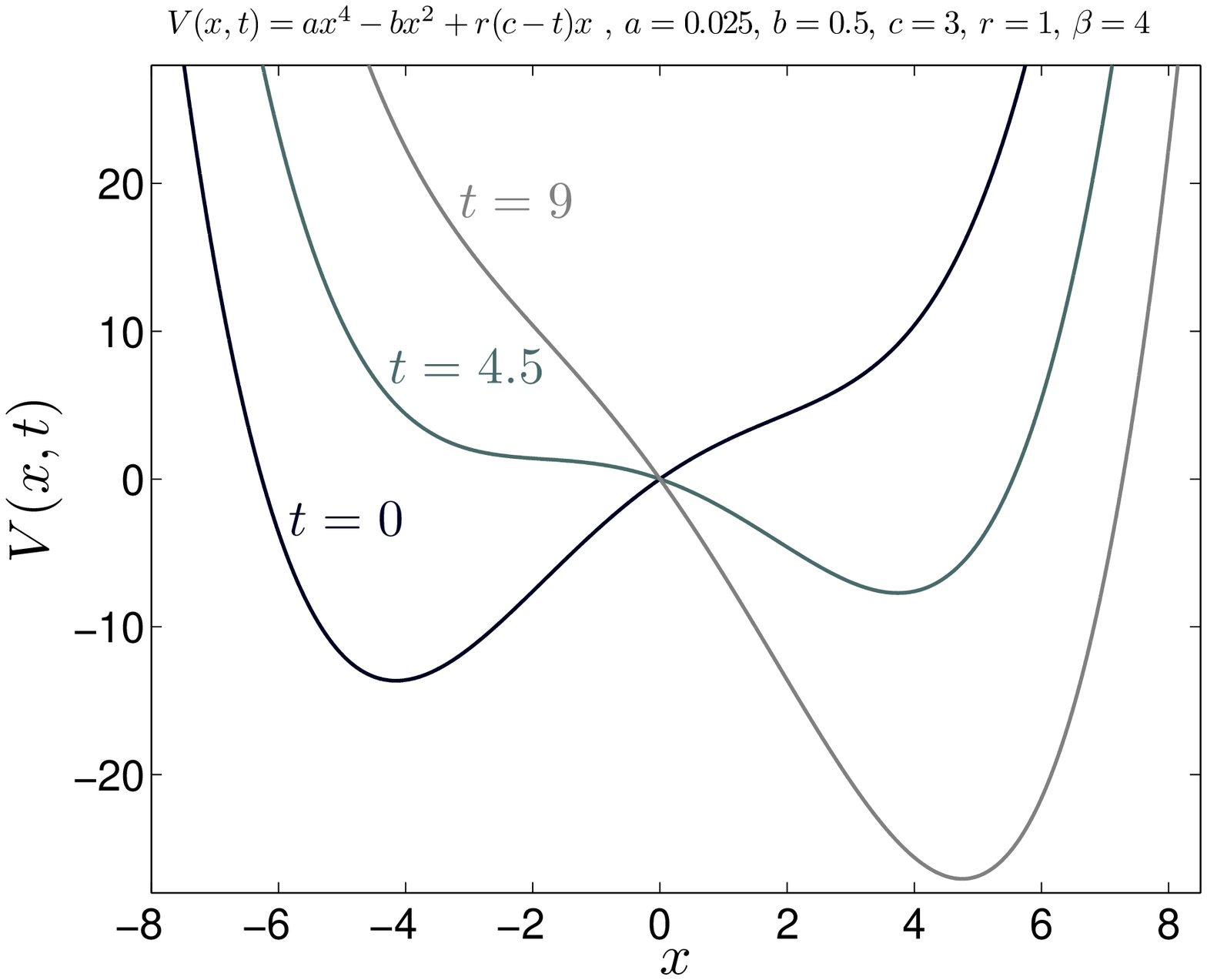}} \\ % 30
  \centering
  \caption{Evolution of the potential $V(x,t)$ (\ref{e:tilsun_V}) in the time-interval $0<t<T$ for two exemplary parameter sets. Shown is $V(x,t)$ for $t=0$, $t=T/2$ and $t=T$ respectively. (a) $a=0.25$, $b=0.5$, $c=2$, $r=3$, $T=6$ and (b) $a=0.025$, $b=0.5$, $c=3$, $r=1$, $T=9$.}
  \label{f:pots}
\end{figure*}

We now apply the analytic method to obtain the asymptotics (\ref{eq:main}) of the work distribution of the dynamics defined by the potential $V(x,t)$ (\ref{e:tilsun_V}). To clarify the procedure, we compile the necessary equations. We have from (\ref{eq:defZ_0}) and (\ref{eq:defN})
\begin{eqnarray}
	\label{e:tilsun_Z0}
  \fl Z_0 = \int\, \d x\; \exp\left[-\beta(ax^4-bx^2+rcx)\right] \\
  \fl \N[x(\cdot)] = \exp\left[6a\int_0^T \di t\; x^2(t)-bT \right] \;.
\end{eqnarray}
The ELE (\ref{eq:ELE}) reads 
\begin{eqnarray}
  \label{e:tilsun_ele}
  \eqalign{
    \fl \ddot{\bx} = 48a^2\bx^5 - 32a\bx^3b + r(c-t)\big[12a\bx^2-2b\big] \\
      + 4b^2\bx + r(1-\i \bq) \;, }
\end{eqnarray}
and its boundary conditions (\ref{eq:elebc}) are of the form
\begin{eqnarray}  
  \label{e:tilsun_elebc} 
  \eqalign{
    0 = \bxt_0 - 4a\bx_0^3+2b\bx_0-rc \;, \\
    0 = \bxt_T + 4a\bx_T^3-2b\bx_T+r(c-T) \;. }
\end{eqnarray}
The constraint (\ref{eq:defwork}) is 
\begin{eqnarray}
  \label{e:tilsun_constr}
  W = - r \int_0^T\!\!\!\d t\;\bx(t;\bq) \; .
\end{eqnarray}
Using a standard relaxation algorithm, the numerically solution of equations (\ref{e:tilsun_ele}) - (\ref{e:tilsun_constr}) for desired work values $W$ yields optimal trajectories $\bx(t)$ depicted exemplarily in figure \ref{f:opttrajs}. 

\begin{figure*}
  \subfloat[][]{\label{sf:opttraj1}
	\includegraphics[trim = 0pt 0pt 0pt 0pt, clip, width=0.45\textwidth]{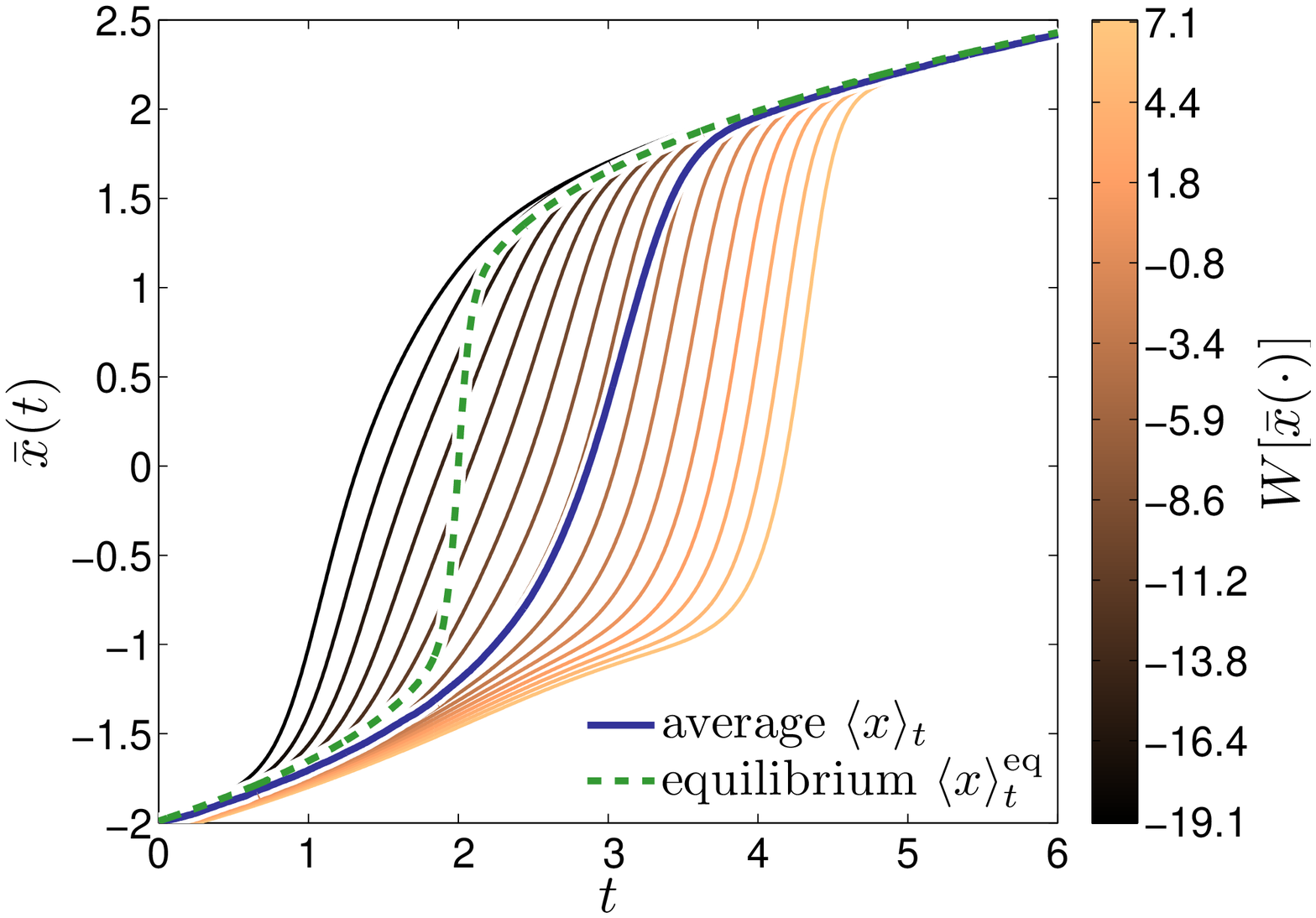}} \qquad % 15
  \subfloat[][]{\label{sf:opttraj2}
	\includegraphics[trim = 0pt 0pt 0pt 0pt, clip, width=0.45\textwidth]{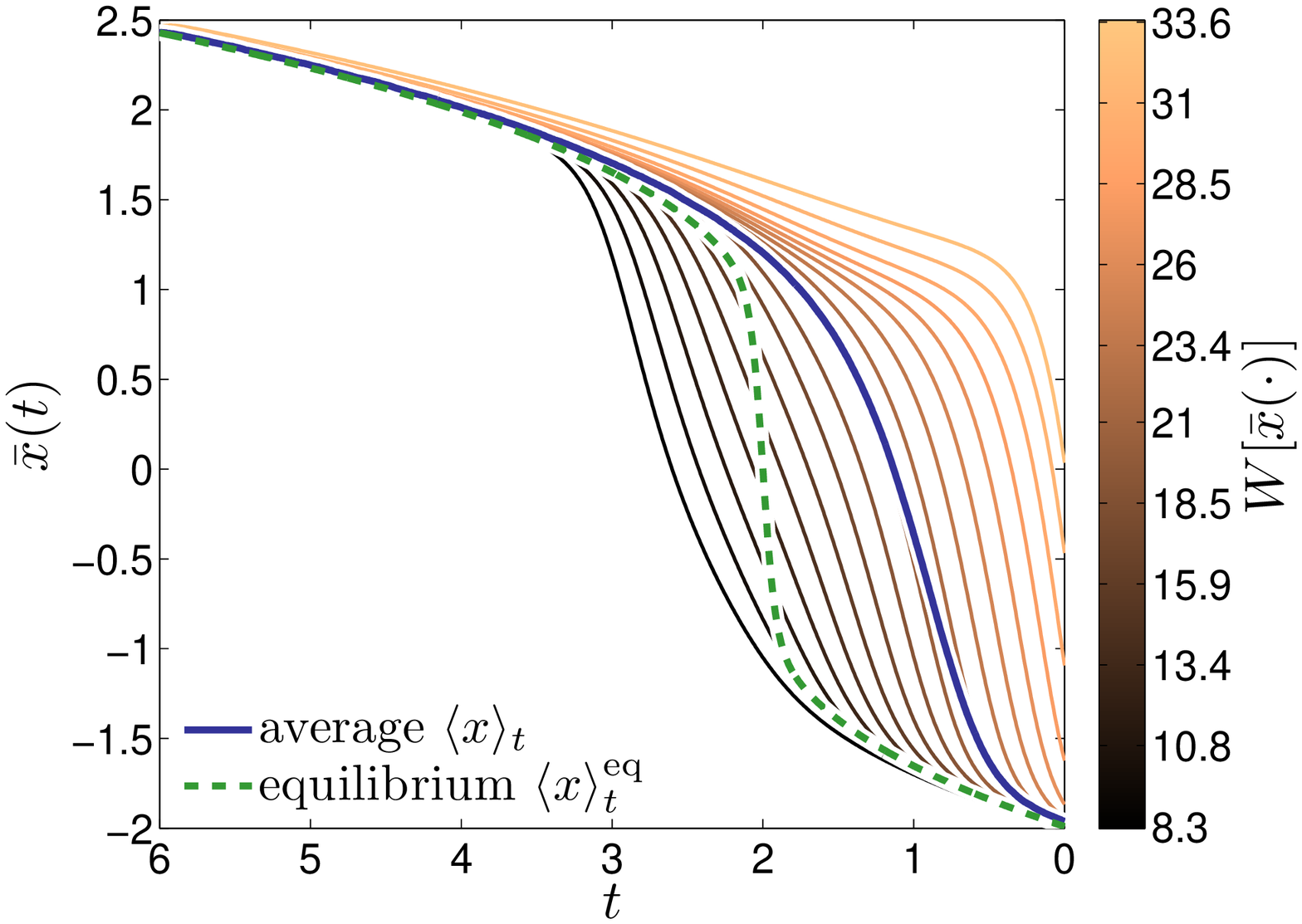}} \\ % 15
%   \subfloat[][]{\label{sf:traj1}
% 	\includegraphics[width=0.45\textwidth]{tilsun2_xfw.eps}} \qquad
%   \subfloat[][]{\label{sf:traj2}
% 	\includegraphics[width=0.45\textwidth]{tilsun2_xrw.eps}} \\ 
  \centering
  \caption{Trajectories for the potential shown in figure \ref{sf:pot1} for (a) the forward and (b) the reverse process. Shown are optimal trajectories $\bx(\cdot)$ for exemplary work values $W$. For a comparison is plotted on top (thick lines), the average trajectory of the simulation (full blue line), and the average $\langle x\rangle_t^{\mathrm{eq}}$ from the equilibrium distribution corresponding to the instantaneous values of the parameters (dashed line).}
  \label{f:opttrajs}
\end{figure*}

The operator $A$ from (\ref{eq:defA}) acquires the form
\begin{eqnarray}
  \label{e:tilsun_A}
  \eqalign{
   \fl A = -\frac{\d^2}{\d t^2} + 240a^2\bx^4 - 96ab\bx^2  \\
      + 24ar(c-t)\bx + 4b^2 }
\end{eqnarray}
with the boundary conditions (\ref{eq:bcA})
\begin{eqnarray}
\eqalign{
  0 = \big[12a\bx_0^2-2b\big]\ph(0)-\dot\ph(0) \;, \\
  0 = \big[12a\bx_T^2-2b\big]\ph(T)+\dot\ph(T) \;.
  }
\end{eqnarray}
To obtain $\det A$ according to (\ref{eq:defF}), we determine
\begin{eqnarray}
  \label{e:sun_F0}
  \det A = \big[24a\bx_T^2-4b\big]\chi(T) + 2 \dot\chi(T)
\end{eqnarray}
by solving numerically the initial value problem (\ref{eq:odedetA})
\begin{eqnarray} 
  \label{e:sun_chi0}
  \eqalign{
  \fl \ddot\chi(t) = \big[240a^2\bx^4 - 96ab\bx^2 \\
    + 24ar(c-t)\bx + 4b^2\big]\chi(t) = 0 \;, \cr
  \fl \dot\chi(0) = 1 \;, \qquad \dot\chi(0) = 12a\bx_0^2-2b \;. }
\end{eqnarray}
This has to be done for each value of $W$ separately by using the appropriate results for  $\bx(t;W)$ and $\bq(W)$.

The last ingredient for the pre-exponential factor is $\bR$ from (\ref{eq:resAm1}). To determine it, we need to solve the boundary value problem (\ref{eq:hode}), (\ref{eq:bcA}) 
\begin{eqnarray}
  \label{e:tilsun_psi}
    \eqalign{
    \fl \ddot\psi(t)=\bigl[240a^2\bx^4 - 96ab\bx^2 \\
      + 24ar(c-t)\bx + 4b^2 \bigr]\psi(t) + r \;, \cr
     \fl 0 = \big[12a\bx_0^2-2b\big]\psi(0)-\dot\psi(0) \;, \\
     \fl 0 = \big[12a\bx_T^2-2b\big]\psi(T)+\dot\psi(T) }
\end{eqnarray}
for each $\bx(t;W)$ and $\bq(W)$ and use the result in (\ref{eq:resAm1})
\begin{eqnarray}
  \label{e:sun_R}
  \bR = -r\int_{0}^{T}\!\!\!\d t\;\psi(t) \;.
\end{eqnarray} 

Plugging the numerical results for $Z_0$, $\bar{\N}$, $\bS$, $\det A$ and $\bR$ into (\ref{eq:main}), we obtain the final result for the asymptotic form of the work distribution. We carried out this program for the two processes characterized by the parameter sets given in the caption of figure \ref{f:pots}, including for both cases the {\em reversed} processes defined by the substitution $t\to (T-t)$. From simulations of the Langevin equation (\ref{eq:LE}), we also obtained from (\ref{eq:defwork}) the corresponding histograms of work distributions. The values of $\beta$ and $T$ are chosen such that most of the trajectories reach the right minimum of $V$ at the end of the forward process, i.e. the molecules are stretched until virtually all of them unfold (cf. figure \ref{f:trajs}). The results for the asymptotics are shown in figure \ref{f:pWs}, together with the outcome of our numerical simulations.

\begin{figure*}
  \subfloat[][]{\label{sf:traj1}
	\includegraphics[trim = 0pt 0pt 0pt 0pt, clip, width=0.45\textwidth]{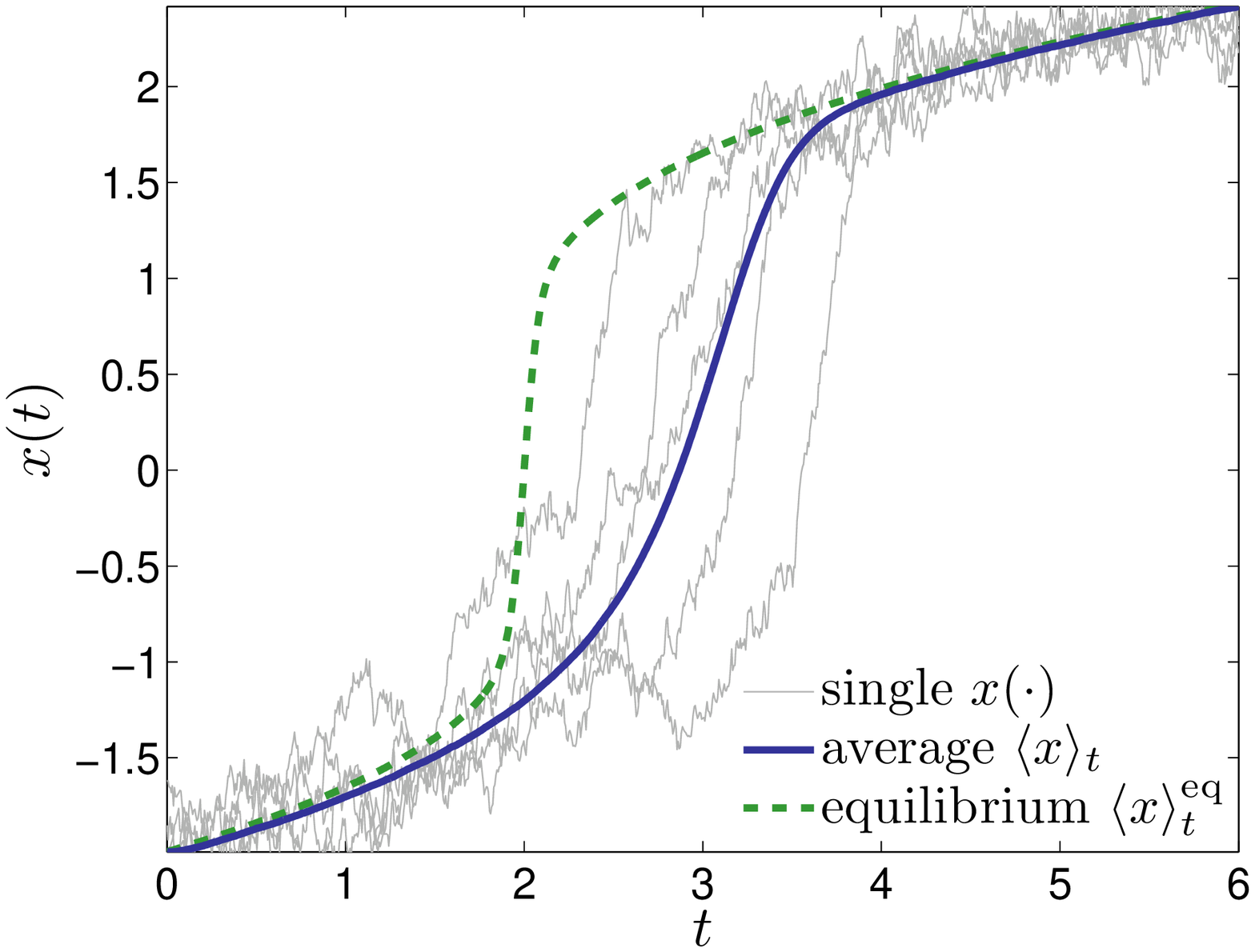}} \qquad % 15
  \subfloat[][]{\label{sf:traj2}
	\includegraphics[trim = 0pt 0pt 0pt 0pt, clip, width=0.45\textwidth]{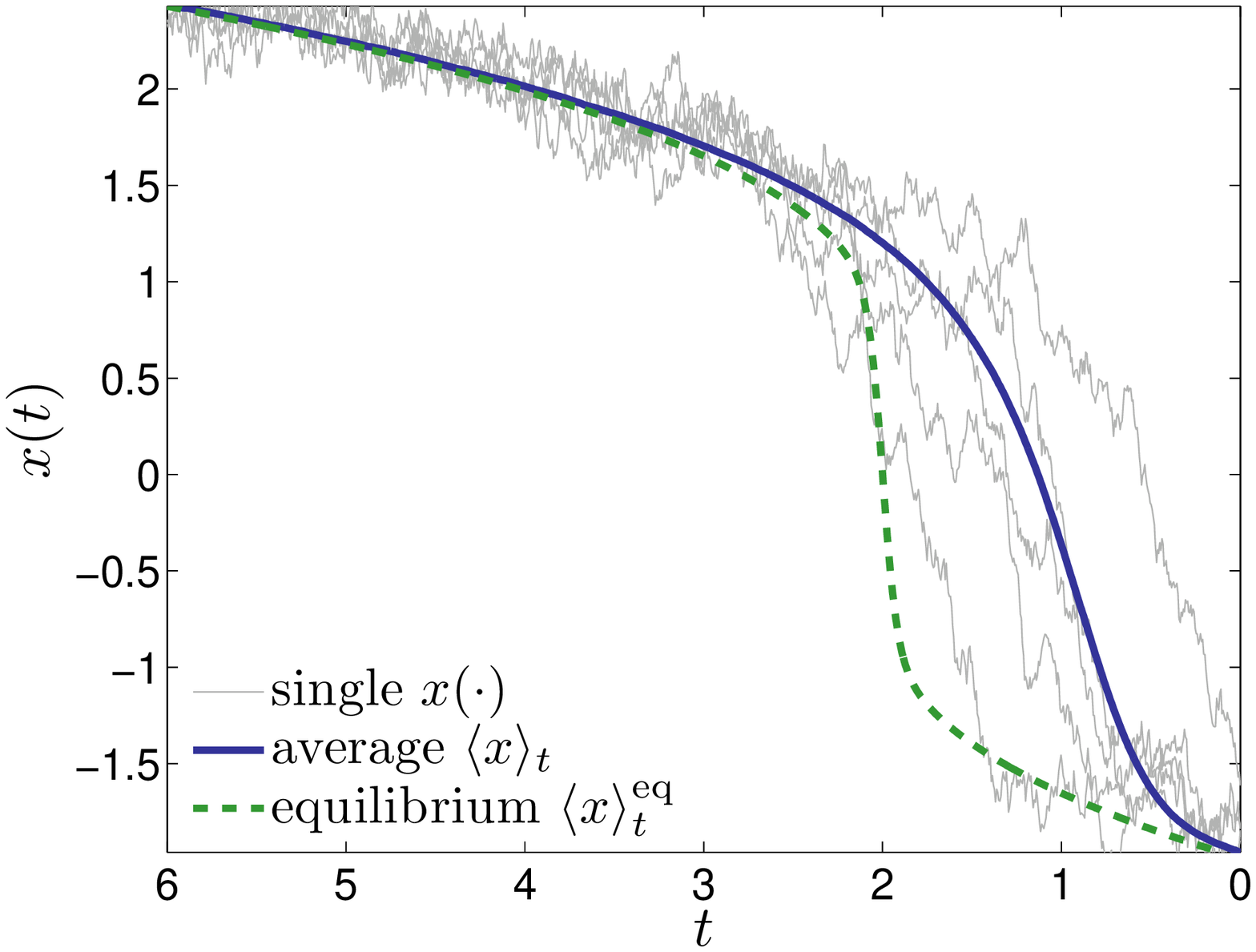}} \\ % 15
%   \subfloat[][]{\label{sf:traj1}
% 	\includegraphics[width=0.45\textwidth]{tilsun2_xfw.eps}} \qquad
%   \subfloat[][]{\label{sf:traj2}
% 	\includegraphics[width=0.45\textwidth]{tilsun2_xrw.eps}} \\ 
  \centering
  \caption{Trajectories for the potential shown in figure \ref{sf:pot1} for (a) the forward and (b) the reverse process. Shown is, exemplified by single realizations $x(\cdot)$, the range of trajectories attainable from the simulation (full grey lines), the average trajectory of the simulation (full blue line), and the average $\langle x\rangle_t^{\mathrm{eq}}$ from the equilibrium distribution corresponding to the instantaneous values of the parameters (dashed line).}
  \label{f:trajs}
\end{figure*}

In a recent paper \cite{Palassini2011}, Palassini and Ritort propose an universal form for the tails of work distributions for single molecule stretching experiments given by
\begin{eqnarray}
  \label{e:pW_fit}
  \fl P(W) \sim n \frac{\Omega^{\al-1}}{|W-W_c|^\al}\,\exp\left[-\frac{|W-W_c|^\delta}{\Omega^\delta}\right] \;.
\end{eqnarray}
Here, $W_c$ is a characteristic work value, $\Omega>0$ measures the tail width, and $n$, $\alpha$ and $\delta$ are further constants. The Jarzynski equation (\ref{eq:JE}) stipulates $\delta>1$.
Based on (\ref{e:pW_fit}), Palassini and Ritort present in \cite{Palassini2011} three slightly different analytical methods to improve the estimation of free-energy differences $\Delta F$ from the Jarzynski equation (\ref{eq:JE}). To decide which approach to use, they distinguish three regimes defined by the parameter combination 
\begin{eqnarray}
  \label{e:lambda_fit}
  \lambda := \left(\delta/\Omega\right)^{\delta/(\delta-1)} \,\ln N \;.
\end{eqnarray}
The three regimes are then identified by $\lambda>1$, $\lambda\ll1$ and $\lambda\lesssim1$ respectively. They test their method with experimental data from DNA stretching experiments. For more details regarding the improved estimation of $\Delta F$ and the experiments see \cite{Palassini2011}.

We fitted the empirical asymptotic form (\ref{e:pW_fit}) to the tails of the work distributions obtained in our simulations by standard least-square fits, starting with a Gaussian distribution specified by $W_c=\lla W \rra$, $\Omega^2=2\lla (W-W_c)^2 \rra$, $\al=0$, $\delta=2$ and $n=1/\sqrt{\pi}$. A subtle point in the procedure is to find the optimal interval of work values for the fit. The resulting parameter values are listed in table \ref{t:fits}, the corresponding fits are included into figure \ref{f:pWs}.

\begin{table}
\caption{\label{t:fits} Fit results for (\ref{e:pW_fit}) for the two examples (a) and (b) shown in figure \ref{f:pots} for the forward (fw) and reverse (rv) process. $N$ denotes the number of realizations used in the simulation, the other parameters are from the proposal (\ref{e:pW_fit}).} 
% (a) FW : Wcut =-2.2037, q =0.00042692, Ω =9.3124,  α =-26.379,  Wc =11.5209, δ =3.3527, µ =1.6186,  λ =4.2199
% (a) RV : Wcut =24.8534, q =1.1155e-05, Ω =8.6398,  α =-31.2481, Wc =38.159,  δ =3.2612, µ =9.2171,  λ =2.2595
% (b) FW : Wcut =8.7747,  q =4.9653,     Ω =8.6056,  α =-3.0146,  Wc =19.9299, δ =2.467,  µ =8.7366,  λ =1.45
% (b) RV : Wcut =30.4109, q =1740.5683,  Ω =26.0791, α =26.526,   Wc =61.1813, δ =3.3278, µ =44.1768, λ =0.60664
\begin{indented}
\lineup
\item[]\begin{tabular}{@{}*{8}{l}}
\br                              
Set     & $N$    & $n$    & $\Omega$ & $\al$ & $W_c$ & $\delta$ & $\lam$ \cr 
\mr
(a) fw  & $10^4$ & 4.27E-4 & 9.31    & -26.4 & 11.5  & 3.35     & 4.22   \cr
(a) rv  & $10^4$ & 1.16E-5 & 8.64    & -31.2 & 38.2  & 3.26     & 2.26   \cr
(b) fw  & $10^5$ & 4.97    & 8.61    & -3.01 & 19.9  & 2.47     & 1.45   \cr
(b) rv  & $10^5$ & 1740    & 26.1    & 26.5  & 61.2  & 3.33     & 0.607  \cr
\br
\end{tabular}
\end{indented}
\end{table}
\begin{figure*}
  \subfloat[][]{\label{sf:pW1}
	\includegraphics[trim = 0pt 0pt 0pt 30pt, clip, width=0.45\textwidth]{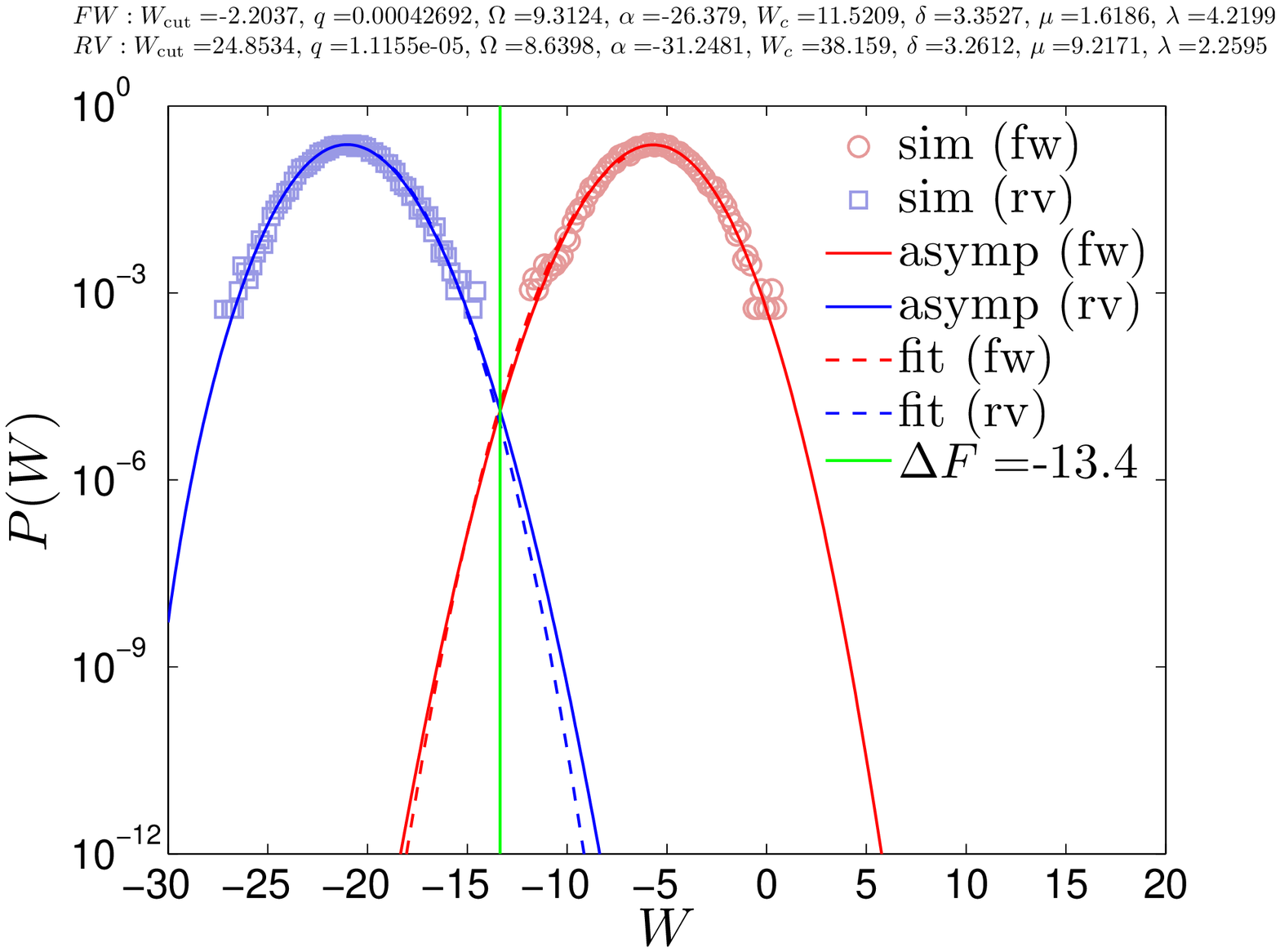}} \qquad % 30
  \subfloat[][]{\label{sf:pW2}
	\includegraphics[trim = 0pt 5pt 0pt 32pt, clip, width=0.45\textwidth]{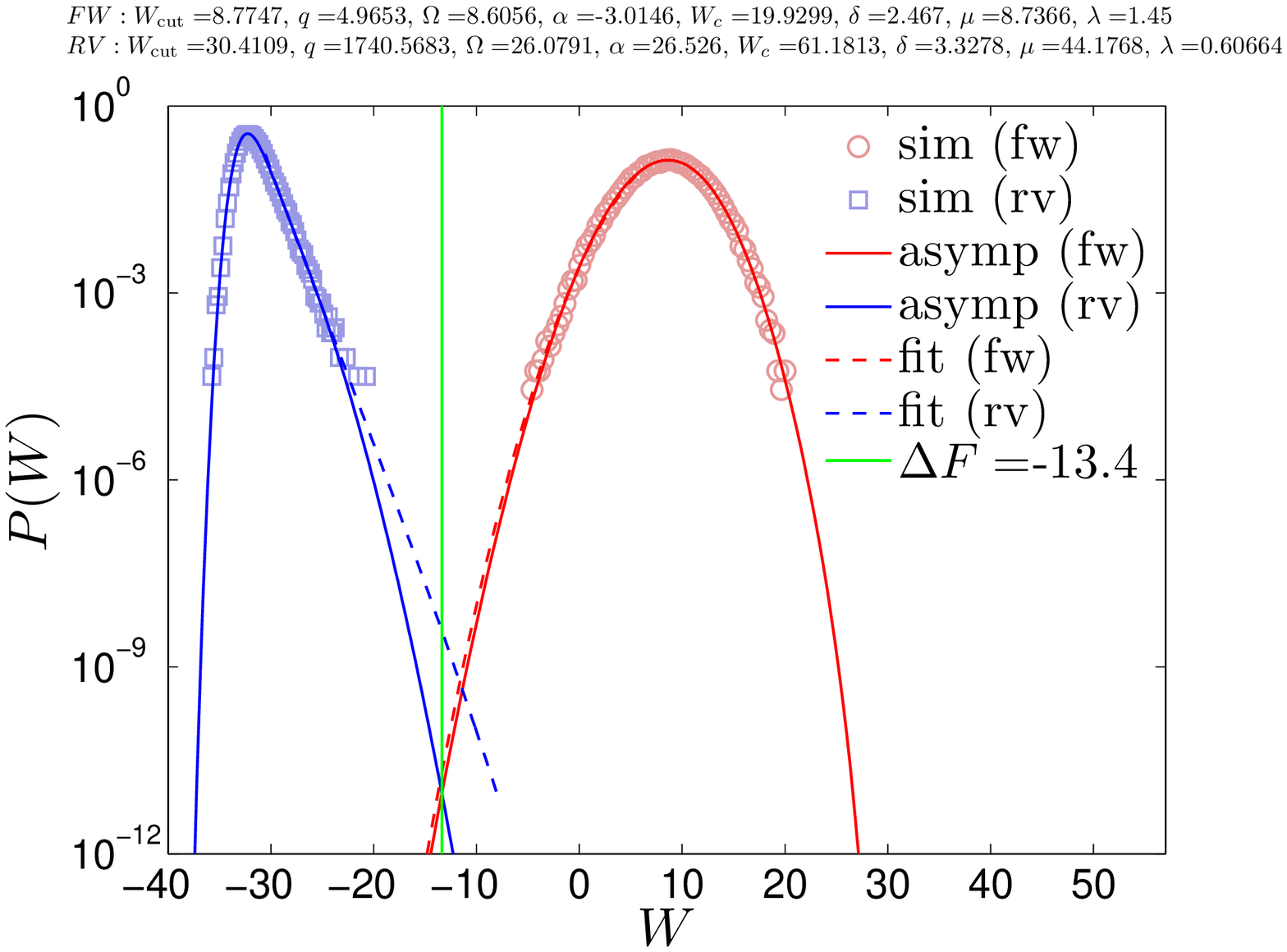}} \\ % 5 0 32 0
  \centering
  \caption{Work distributions $P(W)$ (forward process) and $P_r(-W)$ (reverse process) for the two potentials (a) and (b) shown in figure \ref{f:pots}. Corresponding results obtained from numerical simulations are shown by open circles (forward process) and open squares (reverse process). Full lines depict the asymptotics according to (\ref{eq:main}), dashed lines are fits of (\ref{e:pW_fit}). The value of the free-energy difference $\Delta F$ obtained from numerical integration of the partition functions $Z_0$, $Z_T$ (\ref{e:tilsun_Z0}) is indicated by the vertical line.}
  \label{f:pWs}
\end{figure*}

% \begin{figure*}
%   \subfloat[][]{\label{sf:traj1}
% 	\includegraphics[trim = 0pt 0pt 0pt 0pt, clip, width=0.45\textwidth]{tilsun2_optraj-fw.eps}} \qquad % 15
%   \subfloat[][]{\label{sf:traj2}
% 	\includegraphics[trim = 0pt 0pt 0pt 0pt, clip, width=0.45\textwidth]{tilsun2_optraj-rw.eps}} \\ % 15
% %   \subfloat[][]{\label{sf:traj1}
% % 	\includegraphics[width=0.45\textwidth]{tilsun2_xfw.eps}} \qquad
% %   \subfloat[][]{\label{sf:traj2}
% % 	\includegraphics[width=0.45\textwidth]{tilsun2_xrw.eps}} \\ 
%   \centering
%   \caption{Trajectories for the potential shown in figure \ref{sf:pot2} for (a) the forward and (b) the reverse process. Shown are exemplary optimal trajectories, the average trajectory of the simulation (full blue line), and the average $\langle x\rangle_t^{\mathrm{eq}}$ from the equilibrium distribution corresponding to the instantaneous values of the parameters (dashed line).}
%   \label{f:trajs}
% \end{figure*}

%%%%%%%%%%%%%%%%%%%%%%%%%%%%%%%%%%%%%%%%%%%%%%%%%%%%%%%%%%%%%%%%%%%%%%%%%%%%%%%%%%%%%%%%%%%%%%%%%%%%%%
%%%%%%%%%%%%%%%%%%%%%%%%%%%%%%%%%%%%%%%%%%%%%%%%%%%%%%%%%%%%%%%%%%%%%%%%%%%%%%%%%%%%%%%%%%%%%%%%%%%%%%

\section{Discussion}
% We presented an application of the analytical method proposed in \cite{Nickelsen2011} to obtain the asymptotics of a work distribution. For the potential that defines the dynamics for which the asymptotics is determined, we chose (\ref{e:tilsun_V}) and two parameter sets (a) and (b) (cf. figure \ref{f:pots}) to model the unfolding and refolding of single molecules.
As shown in Figure \ref{f:pWs} for both parameter sets (a) and (b), we achieve an excellent agreement between simulation and asymptotics, not only asymptotically for the tails, but also for the centre of the work distributions. The good reproduction of the centre of the distribution is presumably due to the fact that also the probabilities of {\em typical} work values are dominated by single trajectories and their neighbourhood. Note also the perfect match between the free-energy difference $\Delta F$ and the intersection of our asymptotics $P(W)$ and $P_r(-W)$, which demonstrates that the Crooks relation (\ref{eq:CR}) holds in its exact form for our asymptotics, as has been shown in \cite{Nickelsen2011}.

In addition, figure \ref{f:opttrajs} illustrates the optimal trajectories $\bx(\cdot)$ which dominate the asymptotics of the work distributions (\ref{eq:main}). In comparison with the trajectories obtained from our simulations shown in figure \ref{f:trajs}, the trajectories $\bx(\cdot)$ contributing to the tails of the work distribution are of much broader variety. Interestingly, the probability of small work values is dominated by $\bx(\cdot)$ that run into the evolving right minimum even before the minimum is shaped.

For the parameter set (a), the fit of (\ref{e:pW_fit}) compares well with the analytic asymptotics. In this case a combination of a histogram from experimental values and (\ref{e:pW_fit}) would therefore result in reliable estimates for the free-energy difference $\Delta F$. For the parameter set (b) only the forward process is well described by the fit, for the reverse one the tail of the distribution is markedly overestimated. This shifts the intersection point between $P(W)$ and $P_r(-W)$ away from the correct value of $\Delta F$ as can be seen in figure \ref{sf:pW2}. Also, some fit parameters for this case listed in table \ref{t:fits} clearly deviate from the other cases. To investigate that mismatch more closely, we also fitted (\ref{e:pW_fit}) to the analytic asymptotics (not shown). This results into similar conspicuous fit parameters, but the fitting curve now is almost congruent with the analytic asymptotics. From that we conclude that the empirical asymptotics (\ref{e:pW_fit}) is also valid in this case, but the number of work values obtained from the simulation is not enough to reliably fit (\ref{e:pW_fit}). Note that rather than taking the intersection point between $P(W)$ and $P_r(-W)$, Palassini and Ritort use in \cite{Palassini2011} much more sophisticated methods to obtain estimates of $\Delta F$, based on their empirical asymptotics (\ref{e:pW_fit}). Our investigation aims only at validating (\ref{e:pW_fit}).

In addition to the two parameter sets displayed in figure (\ref{f:pots}), we also investigated several other realizations of the potential (\ref{e:tilsun_V}) (not shown). Mostly, we found that fits of (\ref{e:pW_fit}) to histograms of $10^4$ to $10^5$ work values extrapolate reliably to the asymptotic tail of the distribution. But as exemplified by the case shown in figure \ref{sf:pW2}, there is no guaranty that a number of $10^5$ work values is sufficient.

%%%%%%%%%%%%%%%%%%%%%%%%%%%%%%%%%%%%%%%%%%%%%%%%%%%%%%%%%%%%%%%%%%%%%%%%%%%%%%%%%%%%%%%%%%%%%%%%%%%%%%
%%%%%%%%%%%%%%%%%%%%%%%%%%%%%%%%%%%%%%%%%%%%%%%%%%%%%%%%%%%%%%%%%%%%%%%%%%%%%%%%%%%%%%%%%%%%%%%%%%%%%%

\section{Conclusions}
The asymptotics of work distributions for driven Langevin systems can be determined by the method proposed in \cite{Nickelsen2011}. We employed this method to determine the asymptotics of a system defined by the potential (\ref{e:tilsun_V}) modelling the stretching of single molecules. The unfolding of the molecules corresponds to the forward process, the refolding to the reverse process. We obtained histograms of work by simulating the Langevin equation (\ref{eq:LE}). The form of the work distributions are found to be near-Gaussian, similar to distributions measured in a DNA stretching experiment \cite{Palassini2011}. We observe excellent agreement between asymptotics and work distribution, not only for the asymptotic regime but also for the whole range of work values. 

One aim of single molecule experiments is to obtain the free-energy difference between the folded and unfolded state of the molecule. If both the work distribution for the forward and reverse process is available, the Crooks relation (\ref{eq:CR}) can be used to determine the free-energy difference. It is shown in \cite{Nickelsen2011} that the asymptotics (\ref{eq:main}) generally satisfies the Crooks relation exactly, which we demonstrated for two representative examples of the potential (\ref{e:tilsun_V}).

Finally, we tested the universal form (\ref{e:pW_fit}) of the tails of work distributions proposed by M. Palassini and F. Ritort against our results for the asymptotics (\ref{eq:main}). For a broad range of parameters used for the model potential (\ref{e:tilsun_V}), we found a good agreement between (\ref{e:pW_fit}) and our asymptotics. Only if the work distribution differs markedly from a Gaussian form, a reliable fit of (\ref{e:pW_fit}) is likely to require more data points than usually acquired in single-molecule experiments. This might lead to a significant difference between the exact and estimated free-energy difference as illustrated by our examples.

%%%%%%%%%%%%%%%%%%%%%%%%%%%%%%%%%%%%%%%%%%%%%%%%%%%%%%%%%%%%%%%%%%%%%%%%%%%%%%%%%%%%%%%%%%%%%%%%%%%%%%
%%%%%%%%%%%%%%%%%%%%%%%%%%%%%%%%%%%%%%%%%%%%%%%%%%%%%%%%%%%%%%%%%%%%%%%%%%%%%%%%%%%%%%%%%%%%%%%%%%%%%%

\section*{Acknowledgements} We would like to thank Felix Ritort for interesting discussions and the organizers of the 2011 Nordita workshop "Foundations and Applications of Non-equilibrium Statistical Mechanics" where part of this work was done, for providing a stimulating atmosphere, and the ESF for financially supporting the workshop. D.N. acknowledges financial support from the Deutsche Forschungsgemeinschaft under grant EN 278/7.

\end{document}